\documentclass[fleqn,usenatbib]{mnras}

\usepackage[T1]{fontenc}

\DeclareRobustCommand{\VAN}[3]{#2}
\let\VANthebibliography\thebibliography
\def\thebibliography{\DeclareRobustCommand{\VAN}[3]{##3}\VANthebibliography}

\usepackage{graphicx}	% Including figure files
\usepackage{amsmath}	% Advanced maths commands
\usepackage{amssymb}	% Extra maths symbols
\usepackage{gensymb}
\usepackage[dvipsnames]{xcolor}
\usepackage{newtxtext,newtxmath}

\title[Galaxy merger observability]{The observability of galaxy merger signatures in nearby gas-rich spirals}

\author[Rebecca McElroy et al.]{Rebecca McElroy,$^{1,2,3}$\thanks{E-mail: r.mcelroy@uq.edu.au}
Connor Bottrell,$^{4}$
Maan H.~Hani,$^{5,6}$\thanks{\text{Herschel Fellow}}
Jorge Moreno,$^{7,8}$
Scott M.~Croom,$^{2,3}$\newauthor
Christopher C.~Hayward,$^{9}$
Angela Twum,$^{7}$
Robert Feldmann,$^{10}$
Philip F.~Hopkins,$^{11}$
Lars Hernquist$^{8}$, \newauthor
\& Bernd Husemann$^{12}$
\\
\\
$^{1}$School of Mathematics and Physics, The University of Queensland, St Lucia, QLD 4072, Australia\\
$^{2}$Sydney Institute for Astronomy, School of Physics, Physics Road, The University of Sydney, Darlington NSW, 2006, Australia \\
$^{3}$ASTRO3D: ARC Centre of Excellence for All-sky Astrophysics in 3D\\
$^{4}$Kavli IPMU (WPI), UTIAS, The University of Tokyo, Kashiwa, Chiba 277-8583, Japan\\
$^{5}$Department of Physics \& Astronomy, University of Victoria, Finnerty Road, Victoria, British Columbia, V8P 1A1, Canada\\
$^{6}$Department of Physics and Astronomy, McMaster University, Hamilton, Ontario, L8S 4M1 Canada\\
$^{7}$Department of Physics and Astronomy, Pomona College, Claremont, CA 91711, USA\\
$^{8}$Downing College, University of Cambridge, Cambridge CB3 OHA, UK\\
$^{9}$Center for Computational Astrophysics, Flatiron Institute, 162 Fifth Avenue, New York, NY 10010, USA\\ 
$^{10}$Institute for Computational Science, University of Zurich, Winterhurerstrasse 190, Zurich, Switzerland\\
$^{11}$TAPIR, Mailcode 350-17, California Institute of Technology, Pasadena, CA 91125, USA\\
$^{12}$Max-Planck-Institut für Astronomie, Königstuhl 17, 69117 Heidelberg, Germany\\
}

\date{Accepted XXX. Received YYY; in original form ZZZ}

\pubyear{2022}

\begin{document}
\label{firstpage}
\pagerange{\pageref{firstpage}--\pageref{lastpage}}
\maketitle

% Abstract of the paper
\begin{abstract}

 Galaxy mergers are crucial to understanding galaxy evolution, therefore we must determine their observational signatures to select them from large IFU galaxy samples such as MUSE and SAMI. We employ 24 high-resolution idealised hydrodynamical galaxy merger simulations based on the ``Feedback In Realistic Environment" (FIRE-2) model to determine the observability of mergers to various configurations and stages using synthetic images and velocity maps. Our mergers cover a range of orbital configurations at fixed 1:2.5 stellar mass ratio for two gas rich spirals at low redshift. Morphological and kinematic asymmetries are computed for synthetic images and velocity maps spanning each interaction. We divide the interaction sequence into three: (1) the pair phase; (2) the merging phase; and (3) the post-coalescence phase. We correctly identify mergers between first pericentre passage and 500 Myr after coalescence using kinematic asymmetry with 66\% completeness, depending upon merger phase and the field-of-view of the observation. We detect fewer mergers in the pair phase (40\%) and many more in the merging and post-coalescence phases (97\%). We find that merger detectability decreases with field-of-view, except in retrograde mergers, where centrally concentrated asymmetric kinematic features enhances their detectability. Using a cut-off derived from a combination of photometric and kinematic asymmetry, we increase these detections to 89\% overall, 79\% in pairs, and close to 100\% in the merging and post-coalescent phases. By using this combined asymmetry cut-off we mitigate some of the effects caused by smaller fields-of-view subtended by massively multiplexed integral field spectroscopy programmes.

\end{abstract}

% Select between one and six entries from the list of approved keywords.
% Don't make up new ones.
\begin{keywords}
galaxies:formation -- galaxies:evolution -- galaxies: interactions -- galaxies: kinematics and dynamics
\end{keywords}

%%%%%%%%%%%%%%%%%%%%%%%%%%%%%%%%%%%%%%%%%%%%%%%%%%

%%%%%%%%%%%%%%%%% BODY OF PAPER %%%%%%%%%%%%%%%%%%

\section{Introduction}

Galaxy mergers are integral to the paradigm of hierarchical assembly in the $\Lambda$CDM cosmogony  \citep[e.g.,][]{1978MNRAS.183..341W, 1993MNRAS.262..627L}. In particular, mergers play a vital role in the ex-situ build-up of stellar mass in galaxies and are the main channel for massive galaxies to continue to grow \citep[e.g.,][]{2014MNRAS.444.3986R,2017MNRAS.467.3083R}. Numerical simulations and observational studies demonstrate that merging and recently-merged galaxies also enhance in-situ star formation \citep[e.g.,][]{H89,BH96,MH96,Hopkins2013, Patton2013,2013MNRAS.435.3627E, Moreno2015, 2019MNRAS.482L..55T, Moreno2021}, redistribute gas content \cite[e.g.,][]{2010ApJ...723.1255R, 2012rich, 2012MNRAS.426..549S, Moreno2019}, and increase the incidence of active galactic nuclei \citep[AGN,][]{2011ApJ...743....2S, 2014MNRAS.441.1297S, 2018PASJ...70S..37G,2019MNRAS.487.2491E} relative to carefully-matched non-merging control counterparts. Putting these findings in a cosmological context -- e.g., to determine the relative role of mergers in driving star formation and nuclear activity in galaxies \citep{2010engel, 2013MNRAS.428.2529H, 2013Hung} -- requires a thorough understanding of the observational biases and limitations in merger identification and its sensitivity to merger configurations and stages. 

Identification of mergers is complicated by their diversity and transient nature -- making universally-applicable criteria for merger selection challenging. Although stellar tidal tails, bridges, streams, and shell structures are clear signatures of interactions \cite[e.g.,][]{1972ApJ...178..623T, 1992ARA&A..30..705B, 1993ApJ...417..502H, 1994ApJ...431..604G,1998ApJ...499..112B, 2004AJ....128..163L, 2006AJ....132...71H, 2013MNRAS.429.1051C}, the visibility of these features in images is sensitive to resolution, surface brightness limits, merger configuration, and stage \citep{2004AJ....128..163L, Lotz2008, 2010MNRAS.404..590L, 2010MNRAS.404..575L, 2019ApJ...872...76N, 2019Bottrell,Blumenthal2020, 2020arXiv200500476F}. Techniques such as the Concentration Asymmetry Smoothness (CAS) scheme from \cite{1994ApJ...432...75A} and \cite{2000Conselice} can be used to systematically visually-identify mergers, but in addition to the issues mentioned above - features not exclusive to mergers (such as bars, asymmetric spiral arms, or clumpy HII regions) can also affect the viability of these measurements.

An alternative method is to look for evidence of disruption in the dynamics of a galaxy. Isolated late-type galaxies are observed to have smooth disc-like rotation \citep{2006Ganda}. Secular processes (bars, star formation, winds, spiral arms) are known to produce low-level deviations from circular rotation \citep{2003Conselice} -- mergers and interactions cause much larger deviations due to the changes in the gravitational potential of the system \citep{2015A&A...582A..21B,2015ApJ...803...62H,2017Bloom}. As a result, spatially-resolved kinematics provide an alternative window into the merger state of a galaxy. The advent of large-scale surveys performed with integral field spectrographs (IFS) allows for the gaseous and stellar kinematics of \textit{thousands of galaxies} to be examined. Kinemetry \citep{kinemetry_paper} was developed to measure the deviations from regular rotation in the velocity fields of nearby galaxies observed as part of the Spectroscopic Areal Unit for Research on Optical Nebulae (SAURON) Project \citep{SAURON}. Since its development, this method has been applied to large samples of galaxies, such as the Sydney-Australian-Astronomical-Observatory Multi-object Integral-Field Spectrograph (SAMI) and the Mapping Nearby Galaxies at APO (MaNGA) surveys, amongst others, to identify what fraction of the galaxies are likely to be mergers \citep{2015A&A...582A..21B, 2017Bloom, 2017Jesse, 2020Feng}. These surveys have shown that kinematic asymmetry (the level of deviation from regular rotation) correlates well with visual identification of mergers, and that it also persists after a system has coalesced.  

By applying these kinematic and photometric analyses to a large sample of interacting galaxies we gain insight into how these properties evolve over the course of an encounter by observing the variety of merging phases that galaxies can undergo. However, observations only allow us to see one snapshot in time of a particular system. To quantify how these parameters vary as galaxies interact and merge, we need to employ simulations to access the time domain.

\cite{2016ApJ...816...99H} examined the kinematics of the star-forming gas in binary merger hydrodynamic simulations. They inspected the effect of mass ratio on gas kinematic asymmetry, and found that between 20 and 60 percent of their sample was not detected as mergers in the strong interaction phase (between first passage and coalescence, akin to our merging phase). \cite{2021ApJ...912...45N} fully simulate MaNGA data for snapshots of GADGET-3/SUNRISE simulations of merging galaxies and apply linear discriminant analysis (LDA) of the stellar kinematics to identify mergers. This allows them to correctly identify major mergers with 80\% accuracy. In their previous work, \cite{2019ApJ...872...76N} performed a similar analysis but based on imaging. They combine several of the commonly used image asymmetry parameters and find this much more effective than using any one alone. They conclude that to best leverage the data to detect mergers at all stages and mass ratios they should combine imaging and IFU kinematics \citep{2021ApJ...912...45N}. Conversely, \cite{Bottrell2022} use convolutional neural networks to show identification of merger remnant galaxies in TNG100 only improves marginally when you combine imaging with stellar kinematic data.  

\cite{2016ApJ...816...99H} focus on the gas kinematics, \cite{2021ApJ...912...45N} focus primarily on individual time stamps rather than time evolution, and \cite{Bottrell2022} look only at the merger remnant phase. To build on these important studies we will employ a large suite of high-resolution galaxy merger simulations \citep{Moreno2019,Moreno2021} based on the ``Feedback In Realistic Environments" physics model \citep[FIRE-2][]{Hopkins2018} to examine the photometric and kinematic properties of simulated galaxies at every stage of a merger. These next-generation simulations are sampled to high \textit{spatial} and \textit{temporal} resolution and cover a large set of orbital parameters, which allow us to have enough detail to examine individual snapshots and sufficient time resolution to track the time-evolution of each merging system. In this work we will study the evolution of both \textit{stellar kinematics} and photometric asymmetry over time  by selecting two popular asymmetric indicators, commonly used in observational studies of each, and measure how these vary based on merger state and configuration.

Our central goal is to understand how these kinematic measurements are affected by the merger stage of a given system. By using these idealised simulations, we intend to quantify the fraction of merging time at which we could expect to \textit{detect} that a merger is occurring. We also examine how the size of the field-of-view (FoV), the viewing angle, and the alignment of the discs, affect the detection fraction.

In Section~\ref{simulations} we describe our FIRE-2 merger simulations, and Sections~\ref{sec:dataproc} and \ref{sec:photometry} we describe how we generate the stellar particle velocity maps and synthesised photometric images. Sections~\ref{sec:kinemetry} and~\ref{sec:phot_asym} outline how we determine our measurements of kinematic and photometric asymmetry from the data. In Section~\ref{sec:results} we apply varying fields of view to our simulations, look at the effect of viewing angle and configuration of the mergers, as well as examine the effects  of asymmetry as a function of time

\section{Methods}
\label{sec:methods}

%\subsection{Simulations}

\subsection{Our galaxy merger suite}

Our galaxy merger simulations are based on the FIRE-2 model \citep{FIRE,Hopkins2018}, which employs the meshless finite mass (MFM) mode of the \texttt{GIZMO} hydro solver \citep{gizmo2}.  This framework assumes that star formation occurs in self-gravitating, self-shielding \citep{Krumholz2011} gas denser than 1000 cm$^{-3}$ at 100\% efficiency per local dynamical time. Star formation is regulated by feedback, which includes an approximate treatment of momentum flux from radiation pressure; energy, momentum, mass and metal injection from Type Ia and II SNe, plus mass loss from OB and AGB stars. We employ \textsc{starburst99} \citep{Leitherer1999} to calculate stellar masses, ages, metallicities, feedback event rates, luminosities, energies and mass-loss rates. Our radiative heating and cooling treatment includes free-free, photo-ionisation/recombination, Compton, photoelectric, dust-collisional, cosmic ray, molecular, metal-line and fine-structure processes. 

\cite{Moreno2019} describes our galaxy merger suite in detail \citep[see also][]{2019Bottrell,Moreno2021}.  Initially, the secondary galaxy has stellar mass $=$ 1.2 $\times 10^{10}$ M$_{\odot}$, bulge mass $=$ 7.0 $\times 10^{8}$ M$_{\odot}$, gas mass $=$ 7.0 $\times 10^{9}$ M$_{\odot}$, and halo mass $=$ 3.5 $\times 10^{11}$ M$_{\odot}$. The primary has stellar mass $=$ 3.0 $\times 10^{10}$ M$_{\odot}$, bulge mass $=$ 2.5 $\times 10^{9}$ M$_{\odot}$, gas mass $=$ 8.0 $\times 10^{9}$ M$_{\odot}$, and halo mass $=$ 7.5 $\times 10^{11}$ M$_{\odot}$. \footnote{For more details of the initial conditions of both galaxies please refer to Table 2 of \cite{Moreno2019} and research.pomona.edu/galaxymergers/
isolated-disks-initial-conditions/.}. We follow \cite{Mendel2014} and \cite{Saintonge2016} for our bulge and gas mass choices. We adopt three spin-orbit orientations: near-prograde, near-polar, and near-retrograde, following \cite{Moreno2015}. A range of first pericentric passages are simulated, $\sim$7 kpc, $\sim$16 kpc, and $\sim$27 kpc, in addition to three impact velocities. This results in 27 unique simulation runs, of which only those that coalesce and evolve for 250Myr beyond coalescence. We are then left with 15 unique simulations, which are viewed at 4 viewing angles, resulting in 2700 snapshots. For comparison, we also simulate these aforementioned two galaxies in isolation.

We chose to use this fully-characterised set of idealised (non-cosmological) simulations because we are interested in investigating the effects of certain orbital parameters whilst having full control of other initial parameters. This comes at a cost, as idealised simulations do not fully capture the intrinsic and environmental diversity afforded by cosmological simulations \citep{Moreno2013,Sparre2014,Bustamante2018,Hani2018,Blumenthal2020,Hani2020,Patton2020}. We note that, to some degree, cosmological simulations are also limited by cosmic variance -- especially hydrodynamical simulations, which often sacrifice box size to maximise resolution. On the other hand, by comparing against isolated controls, we can tease out the effects of merging, effectively placing intrinsic and environmental effects as second-order effects \citep{Patton2013}. Additionally, by choosing idealised simulations, we can prioritise high spatial and temporal resolution (1.1 parsec and 5 Myr) which is not feasible in cosmological simulations. 

Throughout this paper we will refer to the galaxy pair simulations as the \textit{interacting} sample and the isolated galaxy simulations as \textit{isolated} sample. We further divide the interacting galaxies into three phases, based on milestones in the merger. First passage is defined as the first minimum in the separation between the two galaxies and coalescence is defined as the last time central black holes of the galaxies are 0.5kpc apart. 
\begin{itemize}
    \item \textit{\textbf{Pair phase:}} between first and second pericentric passage. 
    \item \textit{\textbf{Merging phase:}}  between second pericentric passage and coalescence.
    \item \textit{\textbf{Post-coalescence phase:}} when the last time the two galaxies nuclei are separated by more than 500 pc and thereafter.
\end{itemize}
We note that there can be considerable variation in the second pericentric distance, but that it does not affect the discussion since we are using phases and that the run time post-coalescence is not uniform across the simulations, but that this does not significantly affect our conclusions as none of the systems evolve long enough to reach dynamical equilibrium. 
\label{simulations}

% ---------------------------------------
% ---------------------------------------
% ---------------------------------------
\subsection{Data processing}
\label{sec:maps}
\noindent

We generate synthetic line-of-sight kinematic data cubes for the two galaxies at each snapshot. Stellar particles are deposited onto a 3-D (position-position-velocity) Cartesian cube using a cubic spline kernel \citep{1992ARA&A..30..543M} with a smoothing length enclosing the 32 nearest stellar neighbours. The stellar velocities are measured in the galaxy's frame of reference, and deposited assuming no intrinsic velocity dispersion within each stellar particle. Each cube is centred at the galaxy's potential minimum in the spatial dimensions, and the galaxy's velocity in the velocity dimension. The FoV is 50 kpc with a spatial resolution of 97 pc pixel$^{-1}$, while the velocity domain extends to $\pm 700$ km sec$^{-1}$ with a velocity resolution of 4.6 km sec$^{-1}$ pixel$^{-1}$. 

The kinematic cubes are produced along four lines-of-sight for galaxies in our pair, merging, and post-merger phases. The lines-of-sight are defined by the vertices of a tetrahedron centred at the primary galaxy. Therefore, for each galaxy and snapshot sample, we generate four kinematic cubes. On the other hand, owing to the symmetry of the isolated sample, we generate kinematic cubes along 10 inclinations and 11 position angles. We elect to use a systematically sampled set of inclinations and position angles (0$^\circ$, 10$^\circ$, 20$^\circ$, 30$^\circ$, 40$^\circ$, 45$^\circ$, 50$^\circ$, 60$^\circ$, 70$^\circ$, 80$^\circ$, 90$^\circ$) to ensure the diversity in galaxy inclinations in the control sample. We exclude the 0$^\circ$ inclination case as due to the near-zero rotational velocity which causes nonphysical $v_{\rm asym}$ values, which reduces the sample to 2200 snapshots.

Kinematic data cubes are generated for $45$ snapshots in each merger run. The first $30$ selected snapshots linearly and uniformly sample between $[t_{\mathrm{peri}} - 100$ Myr$,\;t_{\mathrm{coal}})$, where $t_{\mathrm{peri}}$ and $t_{\mathrm{coal}}$ are the times of first-pericentric passage and coalescence, respectively. Following \cite{Moreno2019}, we define $t_{\mathrm{coal}}$ as the last time that the distance between each galaxy's supermassive black hole (SMBH) exceeds $500$ pc. The following 15 snapshots uniformly sample the post-coalescence phase which is defined in each run as  $[t_{\mathrm{coal}},\;\max(t_{\mathrm{last}},\;t_{\mathrm{coal}}+500$ Myr$)]$ with the restriction that the time corresponding to the last snapshot in the run, $t_{\mathrm{last}}$, must be greater than $t_{\mathrm{coal}}+250$ Myr. The rationale for this restriction is that our merger suite was originally designed to probe the interacting phase, which naturally truncates the post-coalescence period for some of our mergers.   Discarding many of these mergers is avoided by relaxing the condition to 250 Myr after coalescence \citep[for further details on snapshot selection, see Section $2.1.3$ of][]{2019Bottrell}. 23 mergers from the original \cite{Moreno2019} suite satisfy the $t_{\mathrm{last}} \geq t_{\mathrm{coal}}+250$ Myr post-coalescence condition. Cubes are also generated for $10$ snapshots in each of the two isolated galaxy runs. The selected snapshots linearly and uniformly sample the full run-time of the isolated galaxy simulations.

The cubes for the interacting and isolated simulations are then used to compute moment maps of the line-of-sight velocity distributions (LOSVD). For each spatial pixel, we calculate the first three moments of the velocity distribution of the star particles. We choose to use the stellar velocities rather than gas here as stars are subject to fewer transient internal galactic processes such as winds, outflows, or bars -- and are instead largely only subject to the gravitational potential in which they reside.

\subsection{Limitations}
\label{limitations}
%% limitations on the outcomes: 

When using observational techniques on simulations it is important to consider the limits to which we can compare our results. To this end, we chose to focus on only a few variables within our simulations rather than fully simulate observed data. In particular, we wanted to examine the effect of FoV size, holding all other factors constant. We do not alter the seeing or depth of the simulated data when changing the FoV. While this does impact the realism of our measurements, it makes the conclusions more widely applicable rather than tailored to specific instruments. 

Though we analyse the simulated kinematics in the same way as kinematics derived from integral field spectroscopic data. However, it is important to note that the kinematics we use are taken directly from the simulations rather than measured from simulated spectra, as is the case in observations. This means that errors associated with these measurements are not considered, in addition to variations due to the spectral resolution of any particular instrument. We also do not consider instrumental throughput effects.

These analyses are performed on data with effectively infinite observing time, therefore it is important to note that differing exposure time would effect these results. In particular, much greater exposure times are required to obtain high quality spectral data than for photometry. This means that outside of simulations getting photometric data sufficient to do this kind of analysis is much easier than getting kinematic data. 

Another limitation of this work is that we are limited to looking at a single pair of progenitors in a non-cosmological environment with no variation in mass ratio or galaxy properties. We do hope to address this in future work by using both cosmological and simulations that take into account some variation in galaxy parameters.

\label{sec:dataproc}

% ---------------------------------------
% ---------------------------------------
% ---------------------------------------

\subsection{Synthetic photometry}
Synthetic images of merging, post-coalescence, and isolated galaxies from the \cite{Moreno2019} merger suite were generated by \cite{2019Bottrell} -- see their Section 2.2.2 for details on the creation of our synthetic photometry products. \emph{Idealised} photometric images were produced using the Monte Carlo dust radiative transfer code SKIRT\footnote{\url{https://skirt.ugent.be}} \citep{SKIRT1,SKIRT2} and the Sloan Digital Sky Survey (SDSS) \emph{gri} filter response curves \citep{2010Doi}. These images account for dust but do not include noise, atmospheric blurring, or any other survey-realistic effects explored in \cite{2019Bottrell}. Stellar light from stars older than 10 Myr is modelled using a \cite{2001Kroupa} initial mass function and \texttt{starburst99} single-age spectral energy distributions (SEDs). Emission from young stellar populations ($<10$ Myr) and surrounding \ion{H}{ii} regions use \texttt{MAPPINGSIII} SEDs \citep{2008Groves}. Dust is not tracked explicitly in the \cite{Moreno2019} simulation suite. A dust attenuation model was adopted in which it is assumed that (1) the dust distribution traces the metal distribution, (2) $30\%$ of the metals are locked into dust particles, and (3) the dust comprises the multi-component mix of graphite grains, silicate grains, and polycyclic aromatic hydrocarbons from \cite{2004Zubko}.

\label{sec:photometry}

\begin{figure*}
    \centering
    \includegraphics[width=0.8\textwidth]{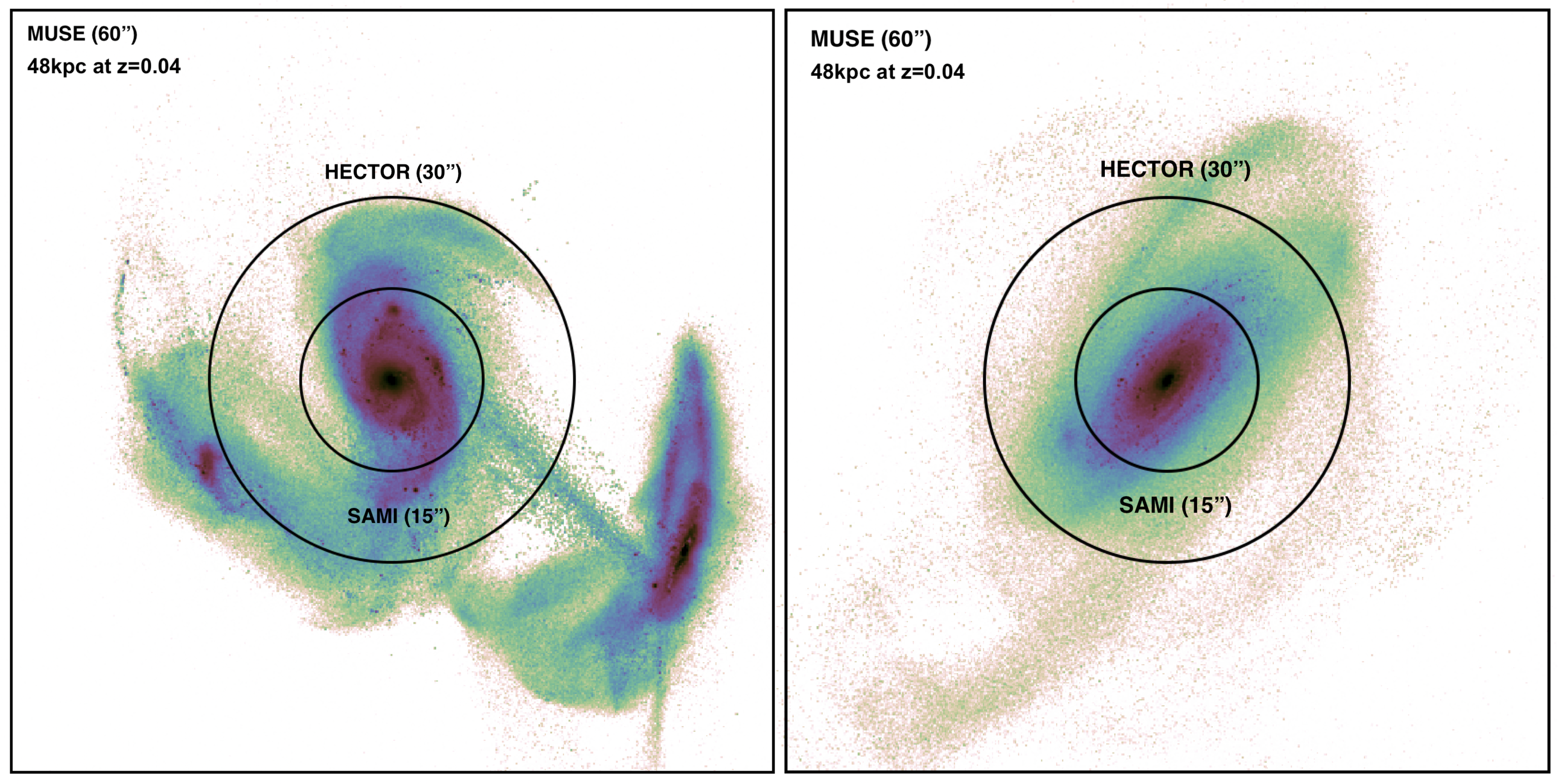}
    \caption{The three FoVs projected onto the simulated photometry of two snapshots from the simulations. The smallest circle shows the SAMI-sized 15" FoV, the larger circle shows the Hector-sized 30" FoV and the full box shows the 60" $\times$ 60" FoV of MUSE. The colour scale represents simulated surface brightness in magnitudes per square arcsecond.}
    \label{fig:FoV}
\end{figure*}

\subsection{Kinemetry}

Triaxial systems, such as galaxies, have kinematic moments that are shown to be highly symmetric, having either {\it even} (point-symmetric, as in the case of surface brightness and velocity dispersion) or {\it odd} (point-asymmetric, as for velocity) parity. When extracted along ellipses,  kinemetry models the first term of odd moments with a cosine law and even moments as constant, assuming regular rotation \citep{kinemetry_paper, Krajnovi__2011}. In much the same way as in photometry, deviations from circular rotation are then determined using higher-order Fourier analysis. This means that, when divided along elliptical rings, the velocity profiles $K(a, \psi)$ can be modelled by equations~1 and 2 from \cite{kinemetry_paper}:

\begin{equation}
K(a, \psi) = A_0(a) + \sum^{N}_{n=1} A_n(a)\sin(n\psi) + B_n(a)\cos(n\psi), 
%(a, \psi) = A_0(a) + \sum^{N}_{n=1} k_n(a)\cos[n(\psi − \phi_n(a))],
\label{kinemetry_harm}
\end{equation}
where
\begin{equation}
\label{modes}
k_n = \sqrt{A^2_n + B^2_n} \qquad \rm{and} \qquad \phi_{\it n} = \arctan(A_{\it n}/B_{\it n}).
\end{equation}
Here, $k_n$ and $\phi_n$ are the amplitude and phase coefficients of the nth term in the harmonic series describing the velocities sampled along an ellipse, $a$ is the semi-major axis length of the ellipse, and $\psi$ is the eccentric anomaly (which, for discs, corresponds to the azimuthal angle measured from the projected major axis in the plane of the galaxy). For a more detailed discussion of the particulars of the kinemetry algorithm, we refer the reader to \cite{kinemetry_paper}. 

By applying kinemetry to our simulated stellar velocity maps we extract the odd harmonic terms, $k_1, k_3$, and $k_5$ along a series of ellipses which are determined by the algorithm. The lowest order term, $k_1$, represents rotational velocity. We choose our ellipses such that they minimise up to the third Fourier component. Therefore, in rotation-dominated systems, such as our simulated galaxies, we expect most of the power to be in the $k_1$ term. The higher order terms that we fit ($k_3$ and $k_5$) represent additional velocity structures on top of the circular rotation, encapsulated by $k_1$, which means that we determine the \textit{kinematic asymmetry} of the velocity map from them. Namely,
\begin{equation}
k_{31} = k_3/k_1 \quad  \textrm{and} \quad k_{51} = k_5/k_1,
\label{k51}
\end{equation}
which, when combined, form
\begin{equation}
v_{\rm{asym}} = \frac{k_3 + k_5}{2k_1}.
\label{asym}
\end{equation}
This $v_{\rm{asym}}$ or kinematic asymmetry is the primary output from the kinemetry analysis and will be used throughout this paper.

\label{sec:kinemetry}

\begin{figure*}
    \centering
    \includegraphics[width=13cm]{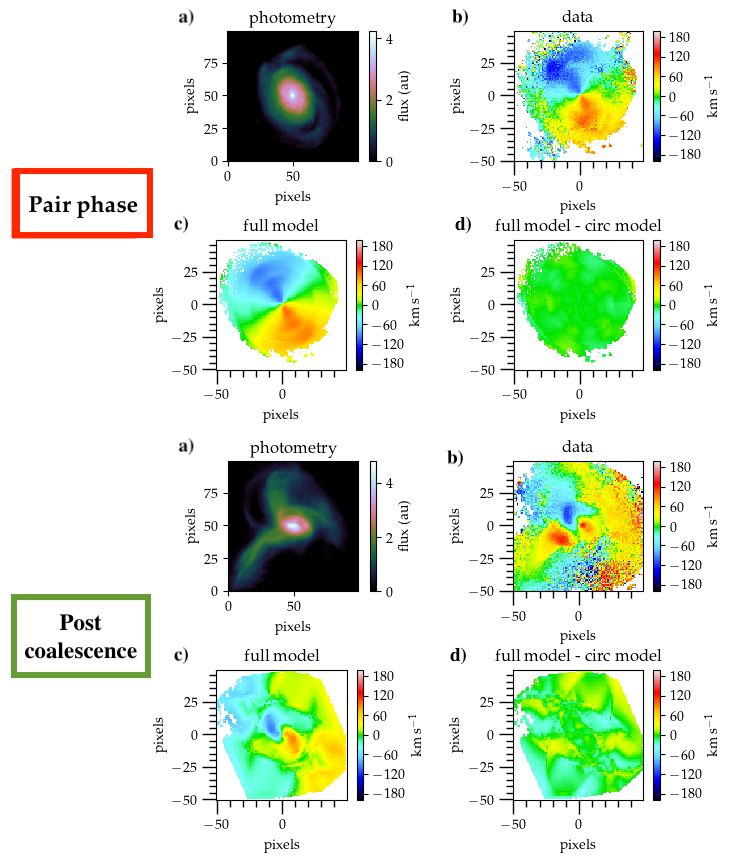}
    \caption{Two examples of the simulated data to which we apply kinemetry. We show a) photometry generated from the simulations, b) the particle velocity data, c) the full kinemetry model, d) the full kinemetry model minus the circular component (i.e. the non-circular components of the model or the deviations from circular rotation). We show these four things for two different snapshots - one after first passage (pair phase)} where kinemetry is still able to model the velocities fairly well and we find low asymmetry. The second shows a snapshot after coalescence where kinemetry struggles to capture the complexity of the stellar particle velocities and we measure high asymmetry.
    \label{fig:kinemetry_ex}
\end{figure*}

\subsection{Photometric asymmetry}

Another standard way to measure the asymmetry of galaxies is to analyse imaging data. Ideally, this might be performed by eye as humans can easily pick out asymmetric or disrupted morphologies at low surface brightness, as demonstrated by large-scale citizen science projects such as Galaxy Zoo \citep[]{2008MNRAS.389.1179L, 2010MNRAS.401.1552D} or large team efforts \citep[e.g. CANDELS;][]{2015Kartaltepe}. Alternatively, several algorithms exist to attempt to measure image asymmetry automatically. The most common of these used in studies of galaxies are the Asymmetry parameter ($A$) from \cite{1994ApJ...432...75A,2003ApJS..147....1C, 2000Conselice} and the M20 \citep{2004AJ....128..163L} and Gini \citep{2003ApJ...588..218A} parameters. We choose to focus on the asymmetry parameter as the complexities of M20 and Gini are beyond the scope of this work. Note, we do not address the use of CNN methods in this work as \cite{2019Bottrell} and subsequent papers investigate this. 

The photometric asymmetry, $A_{\rm{phot}}$, is calculated by rotating an image by 180\textdegree\ and subtracting the rotated image from the original to obtain a residual or difference image. In principle, this residual image should contain all the asymmetric features of the image, and to determine $A$ parameter we take the sum of the absolute value of this residual image, normalised by the sum of the original image. This is given by,
\begin{equation}
    A_{\rm{phot}} = \frac{\Sigma|I_0 - I_{180}|}{2\Sigma|I_0|},
\end{equation}
where $I_0$ is the intensity of a pixel in the original image, and $I_{180}$ is the intensity of that pixel in the rotated image.
\label{sec:phot_asym}

\subsection{Field of view}

One of the goals of this work is to compare how the use of various FoV choices employed by different instruments affects the measured kinematic asymmetry. In order to apply different FoVs to simulations, we need to make some assumptions. The cubes and images both have a fixed FoV of 50 kpc , \citep[using Planck cosmological parameters:][]{2020A&A...641A...6P} and spatial resolution of 97 pc pixel$^{-1}$. Having a similar physical resolution to The Close AGN Reference Survey \citep[CARS,][]{cars_eso} best facilitates future comparisons with our predictions. The median redshift of CARS is $z=0.04$ and, at this redshift, the physical scale is 0.8 kpc arcsec$^{-1}$. The spatial sampling of MUSE in its wide-field mode (WFM) is 0.2" pixel$^{-1}$. If we place the simulations at the median redshift of CARS, $z=0.04$, then each pixel would subtend 0.12". Multiplying 0.12" by the 500 $\times$ 500 pixel simulation box we obtain 60" $\times$ 60" square, which is equivalent to the MUSE WFM FoV. We then re-bin to 100 $\times$ 100 to obtain an angular pixel scale equal to the average seeing for MUSE observations (0.6"). 

Many large galaxy IFS surveys adopt much smaller Integral Field Units (IFUs) than MUSE, meaning that the target galaxies are covered to a smaller radial extent. For example, one of the largest surveys of galaxies, The SAMI Galaxy Survey \citep{sami_dr3}, uses a 15" diameter IFU. The planned upgrade to SAMI, Hector, will feature larger IFUs, with 30" diameter \citep{hector}. 
To test how the radial coverage of galaxies in observations affects the measured asymmetry, we also performed our analysis on smaller fields of view. To do this, we simply placed a 15" and 30" circular aperture onto the simulated data and performed our analysis on the full field and these two smaller fields. It should be noted that these values are typical of and cover the range of FoVs employed by other surveys. CALIFA, like MUSE used a 1' FoV, albeit at much lower spatial resolution \citep{2012A&A...538A...8S}, and MaNGA uses two different FoVs -- 12" and 32" \citep{2015AJ....150...19L}. Figure \ref{fig:FoV} shows two snapshots from the simulations with each of the FoVs labelled. The MUSE-sized FoV is the square on the outside of the images and is the entire simulation box. The two smaller circles represent the SAMI and HECTOR sized FoVs as labelled. We chose two very different snapshots to illustrate how the FoV size will impact different merger stages differently. In the image to the left, many of the merger features (companion, tidal tails) fall outside the two smaller FoVs, rendering them undetectable. In the image to the right, which is the post-coalescence phase, the galaxy is less asymmetric but similarly some tidal features fall outside the smaller FoVs. It should be noted here that we are \textit{only} testing for the effects of FoV - we do not account for the effective resolution (spatially or spectrally).

We work under the assumption that the photometric asymmetry is measured using wide field imaging, and thus we opt to change the size of the field only when calculating the kinematic asymmetry. It would be unrealistic to calculate the photometric asymmetry using the smaller fields because, in observational work, one would almost certainly be using imaging data from one of the large sky surveys such as Galaxy and Mass Assembly (GAMA), The Sloan Digital Sky Survey (SDSS), or the Kilo-Degree Survey (KiDS) \citep{GAMA,2009sdss, KIDS}.

\label{sec:fov}

\section{Results}
\label{sec:results}

\begin{figure*}
    \centering
    \includegraphics[width=0.9\textwidth]{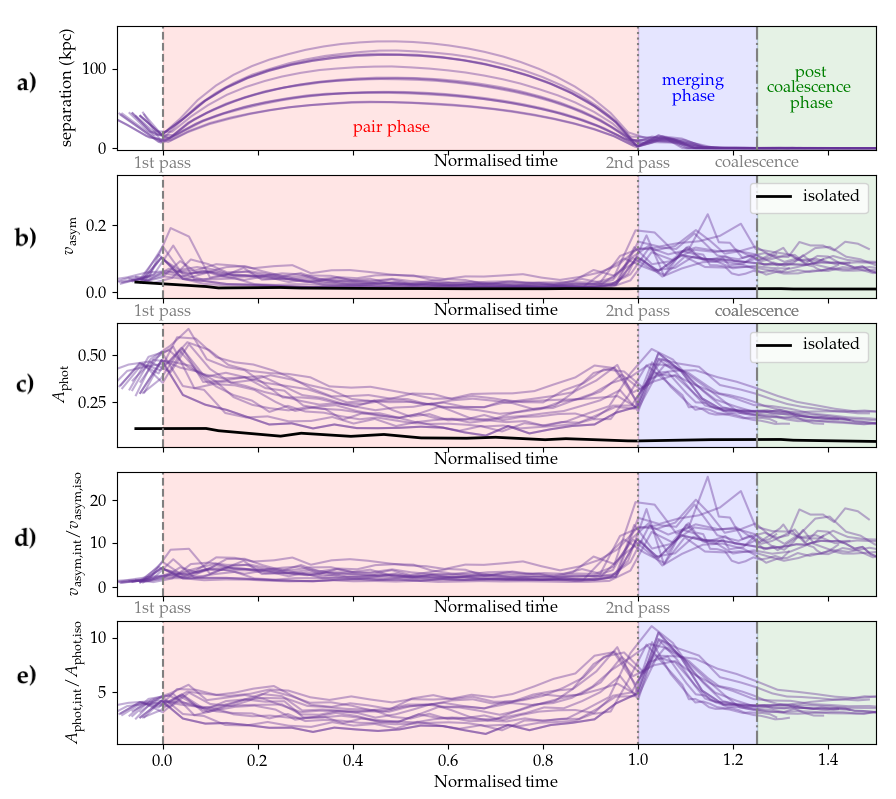}
    \caption{Each line in this figure represents one run of the simulation averaged over viewing angle. a) Separation in kpc versus normalised time in the merger. In this normalised time $t=0$ is first passage (as marked by the dashed line), $t=1$ is second passage (as marked by the dotted line), and the grey dot-dashed line is approximately when coalescence may occur (not precisely as the normalisation is between first and second passages, coalescence occurs the final time the galaxies are separated by 0.5 kpc and varies depending upon the orbital parameters). We also show the three time phases the interacting galaxies go through: the red shaded area is the pair phase when there are two distinct galaxies, the blue shaded area is the merging phase when the galaxies are transitioning from two separate galaxies into one system, and the green shaded area is the post-coalescence phase. b) Kinematic asymmetry vs. normalised time - we see higher asymmetry at first and second passage which is sustained until after coalescence. c) photometric asymmetry vs. normalised time - we see spikes just after first and second passage but with it does not return to as low a level between these times. d) Kinematic asymmetry for the interacting sample divided by the kinematic asymmetry for the isolated sample vs. normalised time. e) photometric asymmetry for the interacting sample divided by the photometric asymmetry for the isolated sample vs. normalised time. 
    }
    \label{fig:all_tracks}
\end{figure*}

Following our analysis of the kinematic and photometric data we evaluate, for every snapshot, a kinematic and photometric asymmetry. This allows us to produce Figure \ref{fig:all_tracks}, which shows separation, kinematic asymmetry, and photometric asymmetry (averaged over viewing angle) against time. The same values for the isolated case are plotted in black. Following \cite{Privon2014}, we normalise time  such that first pericentric passage occurs at $t=0$ and second pericentric passage occurs at $t=1$. We mark these points in time - with a vertical dashed line and a vertical dotted line. 
%See Figures~3 and 4 of \cite{Moreno2019} for the orbital extent of our merger library.
The location of our `coalescence` mark, the grey dash-dotted line, is only approximate because the relative time between second passage and coalescence is not identical for every merger (although these are very similar, as the various blue curves in the light blue box indicate). We can visually represent the three phases that the interacting sample are split into (described in Section~\ref{simulations}) in Figure \ref{fig:all_tracks}. The light red box between first ($t=0$) and second ($t=1$) passage is the pair phase. The light blue box between second passage ($t=0$) and coalescence ($t\sim1.25$) is the merging phase. Finally, the light green box from coalescence to the end of the simulation is the post-coalescence phase. 

Throughout the analysis we focus on the primary galaxy, the more massive of the two, and the data cubes produced from the simulation are centred on the centre of this galaxy. The teal and purple lines in the bottom two panels of Figure \ref{fig:all_tracks} look quite different. This is not unexpected as they are measuring very different quantities. In both $v_{\rm asym}$ and $A$ we see a peak just after first passage. While the galaxies retain their individual structural integrity (i.e. they are not ripped apart and could be described as two separate galaxies) we expect both $v_{\rm asym}$ and $A$ to be high if the galaxies are close together. This is because, in the case of photometric asymmetry - $A$, the image of two superimposed galaxies is asymmetric. It should be noted that the first peak $A$ in is just \textit{after} first passage, which upon examination makes sense, the combined shape of the galaxies will be most asymmetric when they are not maximally superimposed. This is simply because  when maximally superimposed the galaxies will, in our projected view, take up the smallest area of the image. In the case of $v_{\rm asym}$, while the galaxies are separate and retain their own disc rotation but are sufficiently close to one another that they are within the `observed' FoV, \textsc{kinemetry} attempts to model their \textit{combined and superimposed} rotation. It is not surprising then, that the measured kinematic asymmetry - $v_{\rm asym}$ - is high at this point. However, unlike $A$, because at first passage the high $v_{\rm asym}$ is primarily driven by the proximity of the galaxies rather than intrinsic disruption to the kinematics, $v_{\rm asym}$ drops rapidly back down to almost the isolated level when the secondary galaxy leaves the FoV. The photometric asymmetry, $A$, falls when this occurs but not to the isolated level due to the tidal tails left behind by the secondary galaxy. 

Just before second passage $v_{\rm asym}$ rises even higher than at first passage and stays high for much of the rest of the simulation similar to what was found in \cite{Lotz2008, 2015ApJ...803...62H, 2016ApJ...816...99H}, with some reduction further from coalescence. As the galaxies merge the kinematics become very disturbed and chaotic resulting in high measures of kinematic asymmetry, which as time goes on after the merger has occurred will settle down. 
In photometric asymmetry we see a bump just before second passage as the second galaxy re-enters the FoV then a more distinct peak again just after. Unlike $v_{\rm asym}$, the photometric asymmetry falls back to pre-merger levels more quickly. We note that the asymmetries of the isolated sample (black line in bottom panels) do not show significant time evolution and remain fairly constant throughout the simulation.

%\textcolor{red}{text editing required}

\begin{figure*}
    \centering
    \includegraphics[width=0.97\textwidth]{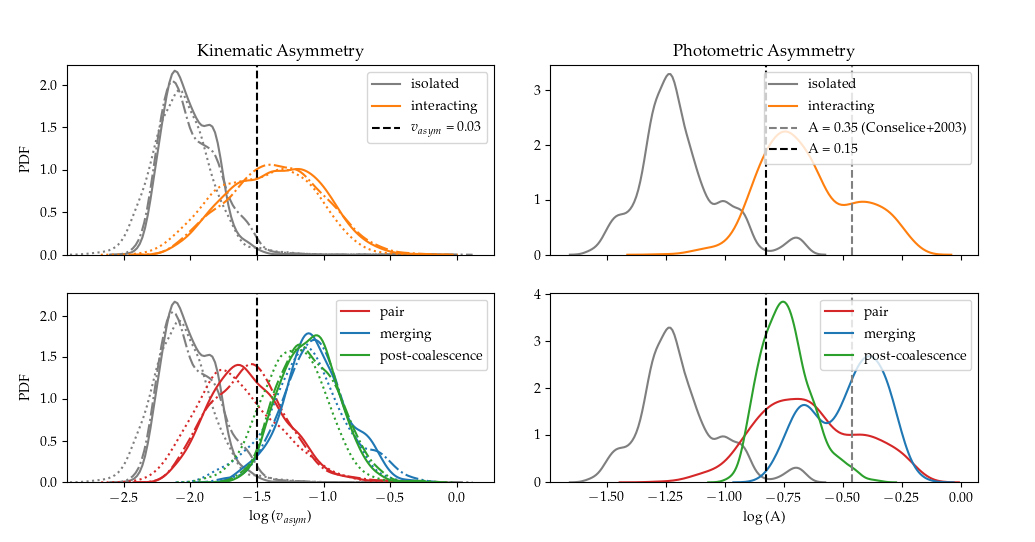}
    \caption{The top row shows the distribution of kinematic and photometric asymmetry for all snapshots in the isolated and interacting samples. As shown in the legend, the isolated galaxies are in grey and the mergers in orange. We also show the $k_{31}$ and $k_{51}$ parameters in addition the combined asymmetry parameter $v_{\text{asym}}$. The dotted line represents the $k_{31}$ parameter, the dash-dotted line represents the $k_{51}$ parameter, and the solid line represents the combined asymmetry parameter. As described in \citet{kinemetry_paper}, $k_3$ is minimised by the fitting and but not $k_5$, 
    but both contain signatures of additional structures. In the bottom row we split the interacting sample up in to the pair, merging, and post-coalescence galaxies. Here, the grey line remains the isolated sample, the red line shows the pair phase, the blue line shows the merging galaxies, and the green line shows the post-coalescence phase. The dashed line in 
    the left column (kinematic asymmetry) shows the 
    cut-off in kinematic asymmetry at 3\%. The grey 
    dashed line in the right column (photometric 
    asymmetry) shows the $A = 0.35$ cut-off from 
    \protect{\citet{2003Conselice}}; we do not adopt this line as 
    it fails to detect most of our interacting sample. We instead adopt the black dashed line, $A = 0.15$.}
    \label{fig:A_asym_4x4}
\end{figure*}

We also compare the results of this analysis for our interacting and isolated samples. Figure \ref{fig:A_asym_4x4} shows a histogram of the measured kinematic and photometric asymmetry for the isolated and interacting galaxies using the full MUSE-sized FoV -- where orange and grey curves denote mergers and isolated galaxies, respectively. Following \cite{2020Feng} and \cite{2017Jesse}, we define a {\it critical} kinematic asymmetry as $v_{\rm asym, \, crit} = 3\%$, above which we would consider a galaxy to be observably `asymmetric' -- in line with judgement ``by eye'' of the regime that excludes isolated galaxies. We display this delineation as a black dashed line. All of the measured asymmetries  from the isolated  sample fall below  this criterion (excepting those from the very beginning of the simulation), and the majority of the interacting galaxies are above it. It should be noted that this delineation is somewhat arbitrary, and moving it in either direction would impact the fraction of galaxies detected as asymmetric. However, we decided that requiring non-detection of the isolated sample as asymmetric was the most logical option. Our cut-off then represents the best case scenario of what could be distinguishable as asymmetric without contaminating the sample with non-mergers.

\begin{figure*}
    \centering
    \includegraphics[width=\textwidth]{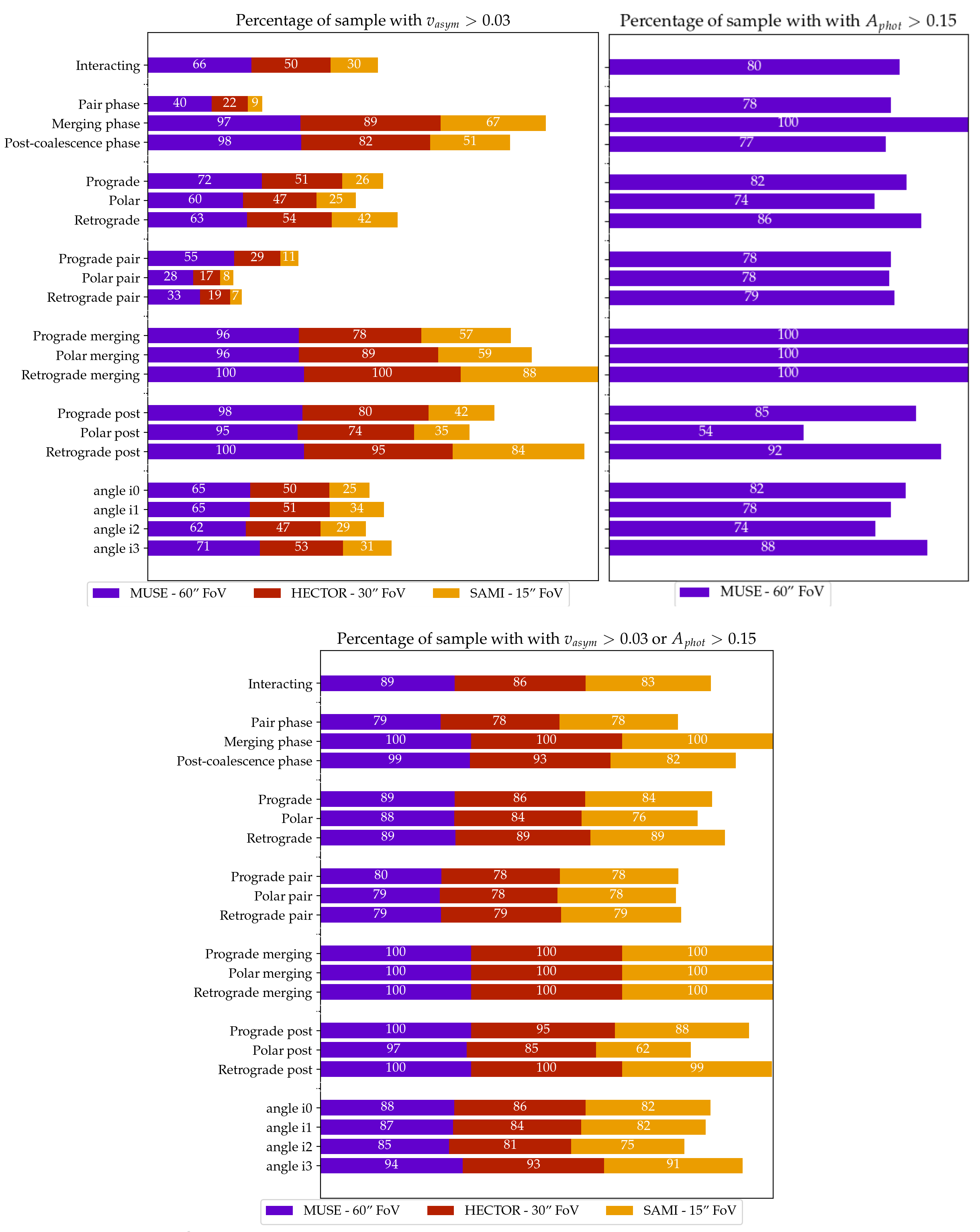}
    \caption{Left: A bar plot showing the fractions of each of the sub-samples that meet the asymmetric criterion of $v_{\rm asym}=0.03$. The different FoVs are shown in three different colours - the full MUSE sized field is purple, the HECTOR sized field is red, and the SAMI sized field is in orange. We see, generally, that as we reduce the FoV the fraction of snapshots meeting the asymmetric criterion falls. We also see that different sub-samples have vastly different detection rates. Right: A bar plot showing the fractions of each of the sub-samples that meet the asymmetric criterion of $A_{\rm phot}=0.15$. Bottom: the two bar plots cut-offs combined.}
    \label{fig:bar_plot}
\end{figure*}

The dynamic range is smaller and the overlap between the two samples is larger in the photometric asymmetry than in the kinematic case. This suggests that photometric asymmetry provides a less definitive measure of disruption than the kinematic one. To address this, we combine both measures in the next section. An alternative way to think about this is to look at the fraction of galaxies that exceed this value in each sample. To show this, we report these fractions in the left of Figure~\ref{fig:bar_plot}, which essentially represent the time spent above the kinematic asymmetry cut-off, or the fraction of the time a merger should be detected given this cut-off. The fraction for each of the sub-samples is shown in each row, with the entire merging sample at the top. We do not show the fraction for the isolated sample because it is approximately 1\% for all fields of view. The right side of Figure \ref{fig:bar_plot} has the same format, but instead uses a cut-off in both the photometric and kinematic asymmetry based on a by-eye delineation between the isolated and merging samples in this plane.

\subsection{Merger stage}
\label{time}

By separating into three phases -- illustrated by the three shaded areas in Figure. \ref{fig:all_tracks} and described in Section~\ref{simulations} -- we can compare the two asymmetry measurements in these different epochs. Recall that the fractions of time that a subset is above the kinematic asymmetry cut-off (for the pair, merging, and post-coalescence phases) are shown in rows 2 to 4 of Figure \ref{fig:bar_plot}. Only 40\% of galaxies in the pair phase satisfy the kinematic asymmetry cut -- including all merger configurations and lines-of-sight. In contrast, galaxies in the merging and post-coalescence phases satisfy the cut with 97\% and 98\% completenesses, respectively. This is likely because, during most of the pair phase, the galaxies are not close together and have only undergone a short interaction that has yet to significantly disrupt the gravitational potential and therefore kinematics. In agreement, \cite{2016MNRAS.461.2589P} also find asymmetry is sensitive to projected separation in galaxy pairs. However, in the merging and post-coalescence phases the galaxies are close together or are merging meaning there will be significant gravitational disruption. The sensitivity in the types of features produced in an interaction to different merger initial conditions (e.g. impact parameter) is explored theoretically in e.g. \cite{1989ApJ...342....1H,1999MNRAS.307..495H, 2008ApJ...689..936J}. Note that the simulations only run for 250-500 Myr after coalescence. This means that the galaxies do not have time to settle into any sort of equilibrium and are therefore still quite irregular. Given more time, it is likely that their asymmetries would fall and a larger fraction would not be detected as mergers. When we use the combined asymmetric cut-off, we increase the number of galaxies that are detected as mergers (82\% as compared to 66\%) -- and particularly in the pair phase (69\% compared to 40\%). The addition of the photometric asymmetry compensates for the loss in IFU FoV, as larger scale asymmetries are still detected in the larger scale imaging - this means that the loss in detectability with FoV size is significantly mitigated.

We can also present this as a histogram with the pair, merging, and post-coalescence phases displayed in Figure \ref{fig:A_asym_4x4} in red, blue, and green respectively. We also show the $k_3/k_1$ parameter in the lightest tone in this figure and the mid-tone line represents the $k_5/k_1$ parameter. We do not identify any significant differences between the $k_3/k_1$ and $k_5/k_1$, so in further plots we will not split the kinematic asymmetry parameter into its component parts. This representation clearly shows the distinct separation between the isolated and merging/post-coalescence phases in kinematic asymmetry, while the pair phase falls in the middle.

\subsection{Viewing angle}
\label{angle}

\begin{figure*}
    \centering
    \includegraphics[width=0.9\textwidth]{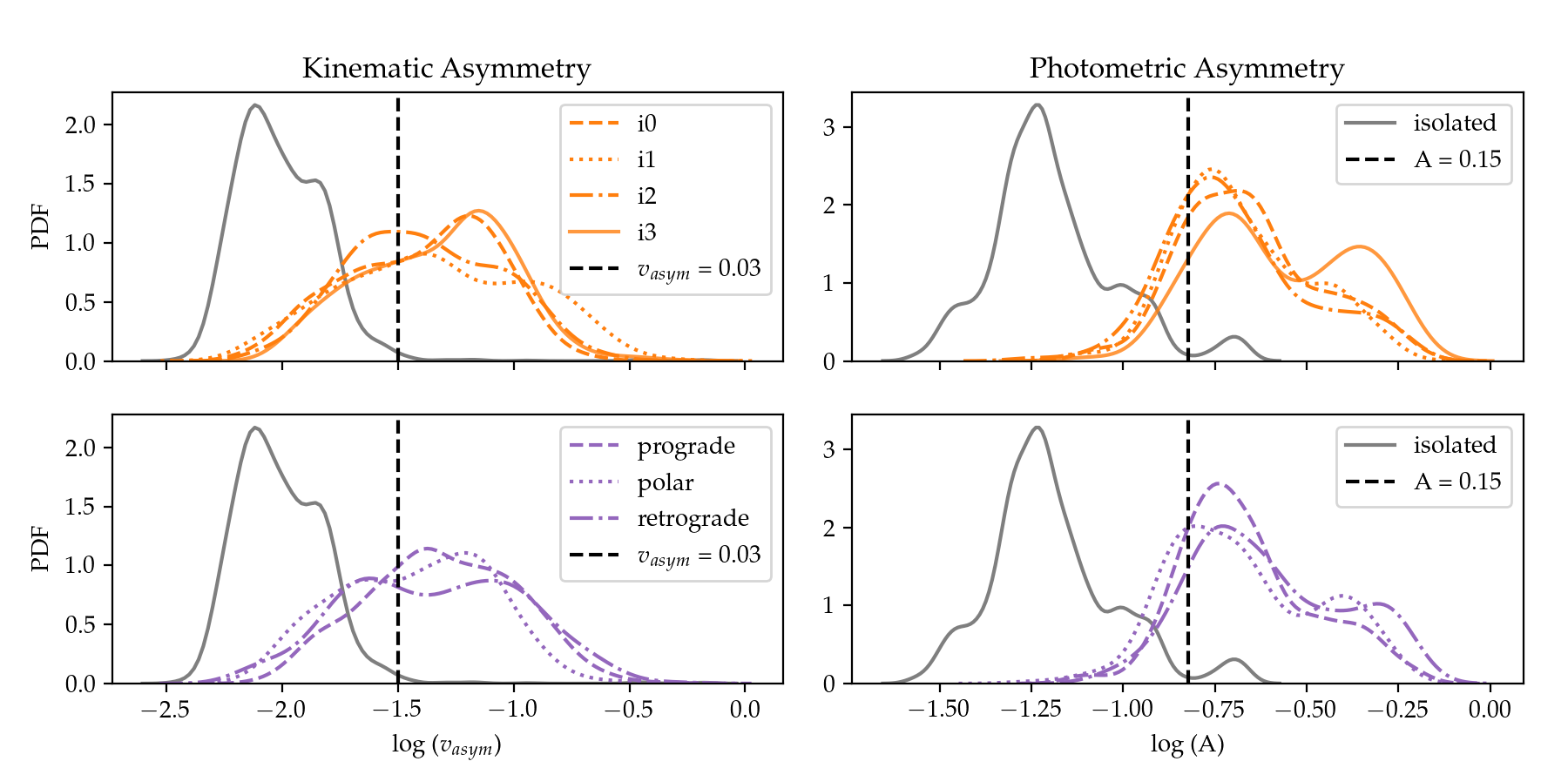}
    \caption{Top row: distribution of asymmetry in the isolated and merging galaxies, with each of the viewing angles shown in a differing line style as shown in the legend. The dashed line shows the cut-off in asymmetry at 3\%. We can see that there is no remarkable difference between the asymmetries measured at various viewing angles. Bottom row: the same, but split into the different configurations. Again, we see no strong difference in the samples. }
    \label{fig:kde_angle}
\end{figure*}

Recall that each snapshot is `observed' at four distinct viewing angles to maximise the size of the sample and increase the variation in the data (Section \ref{sec:photometry}). This also allows us to see whether the measured parameters vary depending on this, and it adds an element of realism -- since real galaxies could be observed at any angle. The top row of Figure \ref{fig:kde_angle} compares the kinematic and photometric asymmetries each of the viewing angles. Rows fourteen through seventeen of Figure \ref{fig:bar_plot} show the fraction of the time the asymmetry measures are found to be above the cut-off.  
We find no significant difference between the angles of observation, except for that i3 (the least face-on angle) shows slightly higher fractions of asymmetry. 

\subsection{Merger configuration}
\label{configurations}

A priori we might expect to witness the greatest asymmetry in the retrograde configuration, since the discs' angular momenta are anti-aligned. The lower panels of Figure \ref{fig:kde_angle} show the distribution of each configuration (prograde, polar, retrograde) compared to the control sample, where we clearly see little difference in the distributions.

\begin{figure*}
    \centering
    \includegraphics[width=0.69\textwidth]{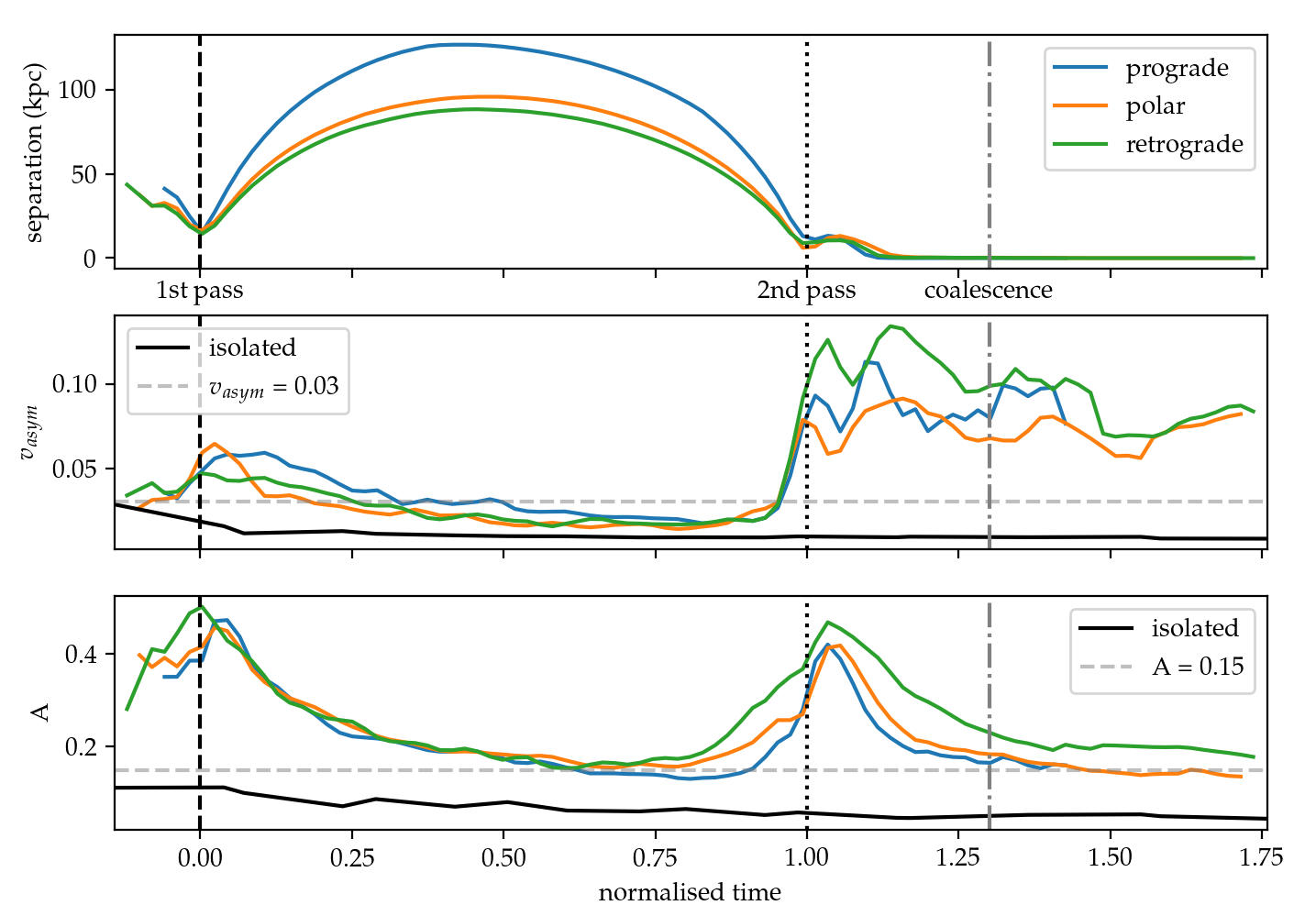}
    \caption{Separation and asymmetry as a function of time for each of the three merger configurations. The top panel shows separation throughout the merger. The middle panel shows kinematic asymmetry with the 3\% asymmetry line marked by the dashed line. The photometric asymmetry is shown in the bottom panel. Each of the configurations is shown in a different colour as indicated by the legend. The average track of the isolated galaxies is shown in black in both lower panels.}
    \label{fig:time_asym_config}
\end{figure*}

We find that while the configurations show little difference averaged over the entire time of the merger -- we do see differences if we bring back the time axis. Figure \ref{fig:time_asym_config} we show tracks of the kinematic asymmetry over normalised time in the merger. Each panel has  normalised time, with first passage, second passage, and coalescence on the x-axis. The top panel shows separation for the three configurations (prograde - blue, polar - orange, retrograde - green) vs. time. The middle panel shows kinematic asymmetry ($v_{\rm{asym}}$) vs. normalised time, with the isolated simulations in black and the kinematic asymmetry cut-off (0.03) as the grey dashed line. The bottom panel shows photometric asymmetry ($A_{\rm{phot}}$). During coalescence the retrograde configurations show significantly higher kinematic asymmetry, and particularly they take much longer to start to fall. 
We also see this if we look in Figure \ref{fig:bar_plot} where the post-coalescence kinematic asymmetric fraction for the retrograde mergers is 1.0 (meaning they all are above the asymmetric criterion). This becomes even more apparent if we focus on the smaller fields of view, where the contrast between the retrograde and the other two configurations becomes stronger. When we inspect the smallest SAMI-sized FoV, 84\% of the coalesced retrograde mergers are kinematically asymmetric compared to 42\% of the prograde mergers and 35\% of the polar mergers. This is because in a retrograde merger the remnant has much more centrally concentrated asymmetry than the other configurations. This means that the asymmetry is still captured even when we adopt a smaller FoV, unlike features such as tails or rings, which will fall outside said FoV. We expect retrograde mergers to be the birthplace of two-sigma galaxies  \citep[galaxies with two, rather than one, velocity dispersion peak;][]{2015Tsatsi} and counter-rotating cores \citep[central regions of the galaxies that are counter rotating as compared to the global rotation;][]{1990ApJ...361..381B} because the combination of retrograde orbits results in a kinematic decoupling in the centre of the resulting galaxy \citep{2015Tsatsi, 2019bryant}.

\subsection{FoV size}
\label{fov}

In these simulations, the 'observed' box is centred upon the central galaxy and extends out to 500 kpc. This means that, at pericentric passages, the satellite galaxy transits into and then out of the observed field. For many galaxy surveys, this is not the case, as they only target a single central galaxy -- and their chosen FoV may only cover this object partially. To investigate how the usage of smaller FoVs affects our measurements, we cut down our full simulation box into two circular apertures, chosen such that these represent the SAMI and HECTOR hexabundles, respectively (from smaller to larger), whilst the full FoV represents MUSE in wide-field mode. Figure \ref{fig:FoV}. shows the relative sizes of these FoVs on a single snapshot. 

We assume the use large-scale imaging survey rather than the IFU data and therefore do not alter the FoV when we measure the photometric asymmetry. In Figure \ref{fig:asym_A_contour}, because the measurement of $A_{\rm{phot}}$ does not change, the y-axis data remains the same across panels, whislt the x-axis (kinematic asymmetry) changes.  
To make the differences between these contours clearer, we add the vertical dashed line at $v_{\rm asym} = 0.03$ for reference.
We note that shrinking the FoV (left to right) diminishes the separation between the three time stages, in addition to lowering the measured kinematic asymmetry. We also see that the overlap with the isolated sample increases as we decrease the FoV. There are several reasons for this to happen. Tidal tails, rings, and irregular morphologies are likely to be invisible if we focus only on the centres of galaxies. This means that only centrally/concentrated asymmetric features will be detected using the smallest FoV, which, while more common in the retrograde mergers - were less prevalent in the other configurations. We touch on this in Section \ref{configurations}, where we highlight that whilst the asymmetric fraction of the prograde and polar configurations drops sharply, when we use the smaller fields of view, this does not happen to the same extent for the retrograde configuration.

\subsection{Kinematic asymmetry vs. photometric asymmetry}
\label{combining}

\begin{figure*}
    \centering
    \includegraphics[width=0.999\textwidth]{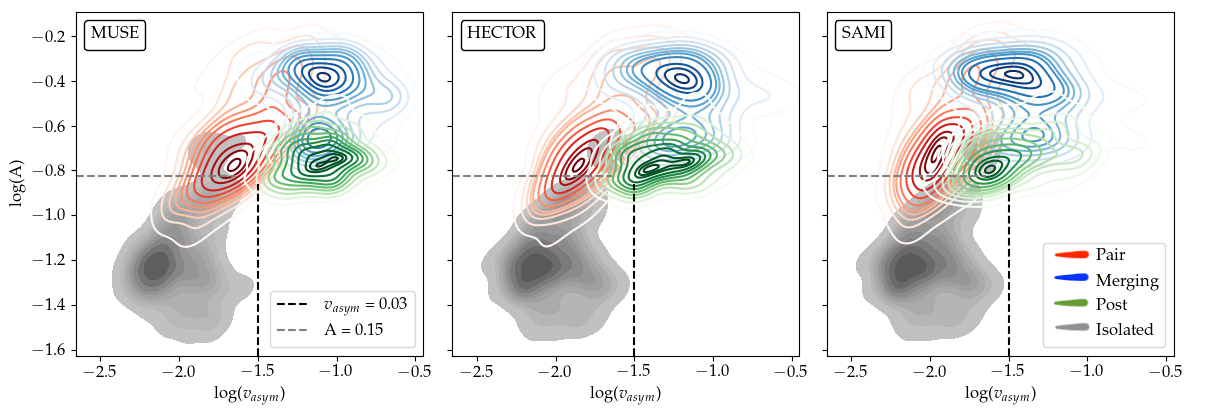}
    \caption{Kinematic vs photometric asymmetry for the three FoV sizes. Here, we have split the data into the three phases - pair (red), merging (blue), and post-coalescence (green). In each panel a contour plot of the distribution of kinematic vs. photometric asymmetry is shown, with the contours 0.1, 0.2, 0.3, 0.4, 0.5, 0.6, 0.7, 0.8, 0.9, 0.95, 0.99. The first panel uses the full FoV (MUSE), the second uses the smaller HECTOR analogue FoV, and the last one uses the smallest SAMI analogue FoV. The vertical black dashed line shows the cut-off in kinematic asymmetry at 0.03, and the horizontal grey dashed line shows the cut-off in photometric asymmetry at 0.15. We take galaxies either line to be detected as asymmetric by our combined asymmetry measure. }
    \label{fig:asym_A_contour}
\end{figure*}

Figure \ref{fig:asym_A_contour} shows the measured kinematic and photometric asymmetry for the pair, merging, and post-coalescence phases. From left-to-right we show the MUSE, HECTOR and SAMI FoV. Grey, red, blue and green contours represent the isolated, pair, merging, and post-coalescence systems.

At the beginning of the pair phase (red), $v_{\rm{asym}}$ and the photometric asymmetry $A$ are high because the two galaxies are either close together or exhibit tidal features. In addition to the greater gravitational disruption, the presence of both galaxies in the field contributes to the measurement. Though this does depend partially on the viewing angle and the orientation of the two galaxies, but on average more of both galaxies are present in the field at this time. When the galaxies are further apart, the gravitational disruption falls and the central galaxy dominates the measurements of asymmetry, resulting in a lower measured value. This is supported by \cite{Moreno2019}, who show that interaction-induced structural effects (within 10 kpc) after first passage only last about 0.5 Gyr. 
Then, before the merging stage, it will increase again due to the proximity of the galaxies. 
The blue contours represent the merging phase. In this stage, the galaxies are close together again as this is just before coalescence. This means the gravitational disturbance is at its greatest as the galaxies merge, resulting in the highest measurements of kinematic asymmetry. In the shorter post-coalescence phase, after the chaos of the merger, the galaxy begins to relax and the kinematics and stellar light become more orderly.

\section{Conclusions}

Similar to \cite{Lotz2008} and as one might predict, we find that the highest asymmetries (both photometric and kinematic) are measured at first passage and just before coalescence. Using kinematic modelling of the particles at each snapshot in the interacting suite we defined a cut-off kinematic asymmetry, $v_{\rm asym} > 0.03$ and a combined asymmetric cut-off above which we describe this merger as detectable. Applying these cut-offs to our whole data set showed that 66\% of the time the interacting simulations would have kinematic asymmetries above this cut-off and 82\% would be above the combined detection limit. We then broke this down into subsamples and looked at how varying the FoV impacted this fraction. The primary results were: 
\begin{itemize}
    \item Mergers, unsurprisingly, are least detectable in the pair phase. Using the largest FoV 40\% were classed as mergers and using the smallest only 9\% were using only kinematic asymmetry. This means that it is very difficult to detect that a galaxy has a companion using stellar kinematics unless they are very both within your FoV as the disturbance is not sufficient to significantly perturb the motions of the stars. However, when we use the combined cut-off this fraction increases to 79\% and only falls to 76\% in the smallest FoV. By combining the photometric asymmetry in addition to the more accurate kinematic asymmetry we are able to mitigate the loss in detections using the smaller FoVs. 
    \item In the merging and post-coalescence phases using the full FoV, almost all (97\% and 98\%) are detected as mergers using both asymmetric cut-offs. However, similarly, when using the smaller FoVs this number can fall as low as 35\% in the post-coalescence phase using a SAMI-sized FoV and only kinematic asymmetries. The combined asymmetric cut-off detects 97\% and 82\% of the merging and post-coalescence galaxies even in the smallest FoV. 
    \item Retrograde mergers show much more persistent kinematic merger features. While the kinematic asymmetry detection fraction for the prograde and polar galaxies falls from 72\% and 60\% in the full FoV to 26\% and 25\% in the smallest FoV - the retrograde mergers only change from 63\% to 42\%. This is even clearer if we look specifically at the smallest FoV observations of the post-coalescence phase where retrograde mergers show at least double the detection fraction compared to the other configurations. This tells us that retrograde mergers are producing centrally concentrated kinematic asymmetries that are just as easy to detect with small FoVs as large ones.
\end{itemize}

While these detection rates are not enormously different to those in \cite{2021ApJ...912...45N} and \cite{2013Hung}, they do surpass them. The particular difference in this study is that we are, due to the fine time sampling of the simulations, able to quantify the detection rates at different stages (pair, merging, post-coalescence) and as a function of time. By doing this we are able to show that our ability to observationally detect that a system is undergoing a merger is highly dependant upon the stage in which we observe it. Additionally, we provide an estimate of how altering the FoV will impact one's ability to detect mergers, which is an important consideration when planning future surveys.

These detectability rates can be used predicatively in present and future large IFU surveys of interacting galaxies using these techniques. In particular, we have shown that when using a smaller FoV and relying only on kinematic information - mergers are significantly under-detected. We also showed that folding in criteria based on imaging helped with this problem and increased detection rates. This means that if future surveys using smaller IFUs wish to detect mergers in their samples, the best strategy will be to combine their kinematic observations with high quality large-scale imaging data.

\section{Future work}

While this work addresses the detection of mergers at fixed mass ratio we have not explored how varying the size of the in-falling companion effects these measurements. We intend to address this in future works with the mass ratio suites of simulations. 

Additionally, given the abundance of IFU observations this method could then be tested on real-world observational data. Now that it is complete, The SAMI Galaxy Survey -- which contains 3000 galaxies with spatially resolved kinematics measurements -- could be a perfect sample to calibrate these asymmetric cut-offs on observational data. SOSIMPLE \citep{2021MNRAS.502.2296D} is a survey of nearby mergers with MUSE and would serve as a perfect comparison sample to the SAMI Survey to test our predictions based on FoV. 

Another aspect we would like to explore in future works are how the presence of a cosmological environment impacts this work - which could be explored by looking at simulations such as Illustris-TNG50 \citep{2019MNRAS.490.3234N}.

\section*{Acknowledgements}

RM acknowledges and pays respect to the Gadigal people of the Eora Nation, upon whose unceded, sovereign, ancestral lands the University of Sydney is built; and the traditional owners of the land on which the University of Queensland is situated, the Turrbal and Jagera people. We pay respects to their Ancestors and descendants, who continue cultural and spiritual connections to Country. 

RM and SC acknowledge the support of the Australian Research Council [Grant ID: DP190102714]. RM also acknowledges the support of a UQ Postdoctoral Fellowship. The computations in this paper were partly run on the Odyssey cluster supported by the FAS Division of Science, Research Computing Group at Harvard University. CB acknowledges support from the Natural Science and Engineering Council of Canada (NSERC) [funding reference number PDF-546234-2020]. MHH gratefully acknowledges support from the William and Caroline Herschel Postdoctoral Fellowship fund. Support for JM is provided by the NSF (AST Award Number 1516374), and by the Harvard Institute for Theory and Computation, through their Visiting Scholars Program. The computations in this paper were run on the Odyssey cluster supported by the FAS Division of Science, Research Computing Group at Harvard University. Support for JM is provided by the NSF (AST Award Number 1516374), the Grace Steele Foundation, and Downing College. RF acknowledges financial support from the Swiss National Science Foundation (grant no 194814). BH greatly appreciates financial support from the DFG via grant GE625/17-1. CCH acknowledges that the Flatiron Institute is supported by the Simons Foundation.

%%%%%%%%%%%%%%%%%%%%%%%%%%%%%%%%%%%%%%%%%%%%%%%%%%
\section*{Data Availability}

 The data underlying this article will be shared on reasonable request
to the corresponding author.

%%%%%%%%%%%%%%%%%%%% REFERENCES %%%%%%%%%%%%%%%%%%

% The best way to enter references is to use BibTeX:

\bibliographystyle{mnras}
\bibliography{bib} % if your bibtex file is called example.bib

% Alternatively you could enter them by hand, like this:
% This method is tedious and prone to error if you have lots of references
%\begin{thebibliography}{99}
%\bibitem[\protect\citeauthoryear{Author}{2012}]{Author2012}
%Author A.~N., 2013, Journal of Improbable Astronomy, 1, 1
%\bibitem[\protect\citeauthoryear{Others}{2013}]{Others2013}
%Others S., 2012, Journal of Interesting Stuff, 17, 198
%\end{thebibliography}

%%%%%%%%%%%%%%%%%%%%%%%%%%%%%%%%%%%%%%%%%%%%%%%%%%

%%%%%%%%%%%%%%%%% APPENDICES %%%%%%%%%%%%%%%%%%%%%

% \appendix

% \section{Some extra material}

% If you want to present additional material which would interrupt the flow of the main paper,
% it can be placed in an Appendix which appears after the list of references.

%%%%%%%%%%%%%%%%%%%%%%%%%%%%%%%%%%%%%%%%%%%%%%%%%%

% Don't change these lines
\bsp	% typesetting comment
\label{lastpage}
\end{document}